\documentstyle[11pt]{article}

\makeatletter
\@addtoreset{equation}{section}
\makeatother

\let\la=\label  
   
\def\nn{\nonumber} \def\bd{\begin{document}} \def\ed{\end{document}}
\def\ds{\documentstyle} \let\fr=\frac \let\bl=\bigl \let\br=\bigr
\let\Br=\Bigr \let\Bl=\Bigl 
\let\bm=\bibitem
\let\na=\nabla
\let\pa=\partial \let\ov=\overline 
\newcommand{\be}{\begin{equation}} 
\newcommand{\ee}{\end{equation}} 
\def\ba{\begin{array}}
\def\ea{\end{array}}
\def\ft#1#2{{\textstyle{{\scriptstyle #1}\over {\scriptstyle #2}}}}
\def\fft#1#2{{#1 \over #2}}
\def\del{\partial}
\def\vp{\varphi}
\def\sst#1{{\scriptscriptstyle #1}}
\def\oneone{\rlap 1\mkern4mu{\rm l}}
\def\td{\tilde}
\def\wtd{\widetilde}
\newcommand{\ho}[1]{$\, ^{#1}$}
\newcommand{\hoch}[1]{$\, ^{#1}$}
\newcommand{\bea}{\begin{eqnarray}} 
\newcommand{\eea}{\end{eqnarray}} 
\newcommand{\ra}{\rightarrow}
\newcommand{\lra}{\longrightarrow}
\newcommand{\Lra}{\Leftrightarrow}
\newcommand{\ap}{\alpha^\prime}
\newcommand{\bp}{\tilde \beta^\prime}
\newcommand{\tr}{{\rm tr} }
\newcommand{\Tr}{{\rm Tr} } 
\newcommand{\NP}{Nucl. Phys. }
\newcommand{\tamphys}{\it Center for Theoretical Physics\\
Texas A\&M University, College Station, Texas 77843}
\newcommand{\ens}{\it Laboratoire de Physique Th\'eorique de l'\'Ecole
Normale Sup\'erieure\hoch{3}\\
24 Rue Lhomond - 75231 Paris CEDEX 05}
\newcommand{\auth}{M. J. Duff\hoch{\dagger1}, H.
L\"u\hoch{\ddagger} and C. N. Pope\hoch{\dagger2}}

\begin{document}

\begin{flushright}
\hfill{CTP-TAMU-18/97}\\
\hfill{LPTENS-97/17}\\
\hfill{hep-th/9704186}\\
\hfill{}
\end{flushright}

\vspace{20pt}

\begin{center}
{ \large {\bf Supersymmetry Without Supersymmetry}}

\vspace{30pt}

\auth

\vspace{15pt}

\hoch{\dagger}{\tamphys}

\vspace{10pt}

\hoch{\ddagger}{\ens}

\vspace{40pt}

\underline{ABSTRACT}
\end{center}

   We present four-dimensional $M$-theory vacua with $N>0$
supersymmetry which, from the perspective of perturbative Type $IIA$
string theory, have $N=0$. Such vacua can appear when the
compactifying $7$-manifold is a $U(1)$ fibration. The missing
superpartners are Dirichlet $0$-branes. Someone unable to detect
Ramond-Ramond charge would thus conclude that these worlds have no
unbroken supersymmetry. In particular, the gravitinos (and also some
of the gauge bosons) are $0$-branes not seen in perturbation theory
but which curiously remain massless however weak the string coupling.

{\vfill\leftline{}\vfill
\vskip	10pt
\footnoterule
{\footnotesize
	\hoch{1} Research supported in part by NSF Grant PHY-9411543.
\vskip	-12pt} \vskip	10pt 
{\footnotesize
	\hoch{2} Research supported in part by DOE 
Grant DE-FG05-91-ER40633. \vskip	-12pt} \vskip 10pt
{\footnotesize
        \hoch{3} Unit\'e Propre du Centre National de la Recherche
Scientifique, associ\'ee \`a l'\'Ecole Normale Sup\'erieure 
\phantom{abcde} et \`a
l'Universit\'e de Paris-Sud.}} 

\pagebreak
\setcounter{page}{1}

\section{Introduction}
\la{Introduction}

Both perturbative and non-perturbative effects of ten-dimensional
superstring theory have now been subsumed by eleven-dimensional
$M$-theory \cite{Howe,Luduality,Hulltownsend,Townsendeleven,%
Wittenvarious,Duffliuminasian,Becker1,
Schwarzpower,Horava,TownsendM,Aharony,DuffM,BanksM}.  In particular,
the $D=10$ Type $IIA$ superstring emerges from $M$-theory compactified
on $S^1$ \cite{Howe, Townsendeleven,Wittenvarious}.  In this picture,
the resulting Kaluza-Klein modes are Dirichlet $0$-branes
\cite{Polchinski} with masses proportional to $1/\lambda$ in the
string metric, where $\lambda$ is the string coupling constant.  They
are thus non-perturbative from the Type $IIA$ perspective.  This may
also be seen from the fact that perturbative string states carry no
Ramond-Ramond $U(1)$ charge whereas the massive Kaluza-Klein modes are
necessarily charged under this $U(1)$.  $M$-theory, on the other hand,
draws no distinction between perturbative and non-perturbative states.
An interesting question, therefore, is whether there is any difference
in the status of {\it supersymmetry} when viewed either from the
perturbative Type $IIA$ string or from the vantage point of
non-perturbative $M$-theory.  In this paper we present
four-dimensional $M$-theory vacua with $N>0$ supersymmetry which, from
the perspective of perturbative Type $IIA$ string theory, have $N=0$.
Such vacua can emerge whenever the compactifying $7$-manifold is a
$U(1)$ bundle over a $6$-manifold.  The missing superpartners are
Dirichlet $0$-branes.  Someone unable to detect Ramond-Ramond charge
would thus conclude that these worlds have no unbroken supersymmetry.
In particular, the gravitinos (and also some of the gauge bosons) are
$0$-branes not seen in perturbation theory but which curiously remain
massless however weak the string coupling.

    The simplest example of this phenomenon is provided by the
maximally-symmetric $S^7$ compactification \cite{Klein} of $D=11$
supergravity \cite{Cremmer} which is presumably also an acceptable
compactification of $M$-theory.  Actually, this is not immediately
obvious. In the literature one encounters two attitudes to $M$-theory
which from the point of view of perturbative string theory may be
called {\it revolutionary} and {\it counter-revolutionary}. The
revolutionary will say that having decided that the fundamental theory
is eleven-dimensional $M$-theory whose low energy limit is $D=11$
supergravity, then {\it any} vacuum of $D=11$ supergravity is an
acceptable vacuum, whether or not it is a vacuum of Type $IIA$
supergravity \footnote{We are assuming throughout that solutions of
Type $IIA$ supergravity may be elevated to solutions of the full Type
$IIA$ string.}.  ($K3$ \cite{Duffnilssonpope3} provides a nice example
of a $D=7$ vacuum that follows from $D=11$ but not from Type $IIA$
supergravity.)  The counter-revolutionary, on the other hand will
argue that $M$-theory is nothing more than the strong coupling limit
of the Type $IIA$ string and hence the only acceptable vacua are those
that are also solutions of Type $IIA$ supergravity\footnote{ Similar
remarks apply, {\it mutatis mutandis}, to the heterotic string and
$M$-theory on $S^{1}¥/Z_{2}¥$ \cite{Horava}.}. If true, one
might be tempted to conclude that the compactifying $7$-manifold
${\cal M}_7$ must necessarily be of the form ${\cal M}_6 \times S^1 $.
See, for example, \cite{Ovrut}. However, this turns out to be too
strong a requirement even for the conservative. As we shall see, we
can satisfy both camps if ${\cal M}_7$ is a $U(1)$
fibration\footnote{This may come as no surprise to $F$-theorists.}.
Fortunately, $S^7$ may be regarded as a $U(1)$ bundle over $CP^3$ and
is in fact but one example of a whole class of compactifications of
$D=11$ supergravity corresponding to Hopf fibrations which therefore
also admit the interpretation as vacua of the Type $IIA$ theory
\cite{Nilssonpope}.

      Another reason for trusting that $AdS_{4}¥\times S^{7}$ is an
acceptable vacuum of $M$-theory is that it is simply the $M$-theory
membrane solution \cite{Duffstelle} in the limiting case as we
approach the horizon \cite{Gibbonstownsend,Gibbonsdufftownsend}.

      In \cite{Howe} the worldsheet action of the $D=10$ Type $IIA$
superstring \cite{Green} was derived from the worldvolume action of
the $D=11$ supermembrane \cite{Bergshoeff} by identifying the third
worldvolume coordinate with the eleventh spacetime coordinate. The
same procedure continues to apply if this eleventh coordinate
corresponds to a $U(1)$ fiber instead of a circle. Now, however, the
membrane worldvolume would itself correspond to a $U(1)$ bundle over
the two dimensional worldsheet of the string, rather than a direct
product.

     We complete this introduction with a caveat: when discussing the
four-dimensional Type $IIA$ string spectrum we shall not be interested
in magnetically charged states so we shall use the words
``perturbative'' or ``non-perturbative'' to mean with or without the
inclusion of electrically charged Dirichlet $0$-branes. These
four-dimensional $0$-branes have their ten-dimensional origin in the
massive Kaluza-Klein spectrum which is absent in the perturbative Type
$IIA$ string. The remaining states in the spectrum have their origin
in the massless modes in $D=10$.  However, even this sector will
display certain non-perturbative features peculiar to the $AdS_{4}¥
\times {\cal M}_{6}¥$ backgrounds. We shall return to this issue in
section \ref{versus}.

\section{$U(1)$ fibrations}

It has long been known that Type $IIA$ supergravity can be obtained by
the dimensional reduction of eleven-dimensional supergravity on a
circle \cite{Campbellwest,Huq}.  In particular, this means that any
solution of eleven-dimensional supergravity of the form ${\cal
M}_{{10}}¥\times S^1$ can be re-interpreted as a solution of the
Type $IIA$ theory.  In fact the class of eleven-dimensional solutions
that can be re-interpreted as solutions of the Type $IIA$ theory is
much wider than these.  To be precise, it is not necessary that the
eleven-dimensional solution have a direct-product structure, and in
fact any solution that has the form of a $U(1)$ bundle over some
ten-dimensional base manifold also admits a ten-dimensional
interpretation \cite{Nilssonpope}.  For example, this includes the
case of the $AdS_4\times S^7$ solution of eleven-dimensional
supergravity, since $S^7$ can be written as a $U(1)$ bundle over
$CP^3$.  In configurations of this kind where the bundle is
non-trivial, the Kaluza-Klein vector potential arising from the
reduction from $D=11$ to $D=10$ has a topologically non-trivial form,
analogous to the potential for a magnetic monopole.  Note, by the way,
that there is no circle which cannot be shrunk to a point in $S^{7}$,
in contrast to the direct product $CP^{3}¥ \times S^{1}¥$.  In
this section, we shall review the structure of such bundle solutions
from the eleven and ten-dimensional viewpoints; further details may be
found in \cite{Nilssonpope}.

     We may begin by considering the bosonic sector of eleven-dimensional
supergravity, 
\be
\hat{\cal L} = {\hat e}\, {\hat R} -\ft1{48}\,{\hat e}\, 
{\hat F_4}^2 -\ft16 \ast(d{\hat A}_3\wedge d{\hat A}_3\wedge
{\hat A}_3) \ .\label{d11lag}
\ee
After dimensional reduction to $D=10$ according to the Kaluza-Klein 
prescription
\bea
d{\hat s}^2 &=& e^{-\ft23\phi}\, ds^2 + e^{\ft43\phi}\, (dz+{\cal
A})^2 \ ,\label{metred}\\
{\hat A}_3(x,z) &=& A_3(x) + A_2(x)\wedge dz\ ,\label{ared}
\eea
where $ds^2$ denotes the ten-dimensional metric in the string
frame, and all ten-dimensional quantities are taken to be independent of the
coordinate $z$ on the compactifying circle, we obtain the bosonic
sector of the Type $IIA$ theory, with the Lagrangian
\be
{\cal L} = e\, e^{-2\phi}\,\Big( R +4\, (\del\phi)^2 
-\ft1{12}\, F_3^2\Big) -\ft14 e\, {\cal F}^2  -\ft1{48}\,e\, F_4^2 -
\ft12 \ast(dA_3\wedge dA_3\wedge A_2)\ ,\label{d10lag}
\ee
where ${\cal F}=d{\cal A}$, $F_3=dA_2$ and $F_4=dA_3 -dA_2\wedge{\cal A}$. 
The graviton, dilaton and $2$-form potential originate in the 
Neveu-Schwarz-Neveu-Schwarz sector of the string theory, while the
$1$-form and $3$-form potentials come from the Ramond-Ramond sector. 

     The equations of motion for the Type $IIA$ theory admit solutions in
which the ten-dimensional metric $ds^2$ is a direct sum $ds^2= ds_4^2
+ ds_6^2$ of a four-dimensional anti de-Sitter spacetime $ds_4^2$ and
a six-dimensional space ${\cal M}_6$ with metric $ds_6^2$
\cite{Nilssonpope} that satisfies certain conditions given below.
Substituting into the equations of motion following from
(\ref{d10lag}), we easily see that there exist solutions with
\bea
e^{\phi}&=&\lambda\ ,\qquad F_{\mu\nu\rho\sigma} = \fft{6m}{\lambda}\,
\epsilon_{\mu\nu\rho\sigma} \ ,\qquad {\cal F}_{mn}=\fft{2\mu}{\lambda}\,
K_{mn}\ ,\nn\\ R_{\mu\nu} &=& -12 m^2\, g_{\mu\nu}\ ,\qquad R_{mn}=
2m^2 (K^2_{mn}+3g_{mn})\ ,\label{d10sol} \\
m^2 &=& \mu^2 ,\nn
\eea
where $m$ and $\mu$ are constants, $\lambda$ is the ten-dimensional
string coupling constant, $\epsilon_{\mu\nu\rho\sigma}$ is the
Levi-Civita tensor on $AdS_4$, and $K_{mn}$ is an harmonic 2-form on
${\cal M}_6$ with constant magnitude $K^2=6$. A particularly simple
example of an ${\cal M}_6$ that satisfies the necessary conditions is
when it is an Einstein-K\"ahler space of positive curvature, with
$K_{mn}$ taken to be the K\"ahler form $J_{mn}$, in which case we see
that
\be
R_{mn}=8 m^2\,g_{mn}.
\ee
Since the dimensional reduction from (\ref{d11lag}) to (\ref{d10lag})
is a consistent one, the solution (\ref{d10sol}) may be lifted back to
a solution of the eleven-dimensional theory, with the metric given by
(\ref{metred}):
\be
d{\hat s}^2 = d\hat s_4^2 + d\hat s_7^2\ ,\label{d11sol}
\ee
where $d\hat s_4^2= \lambda^{-2/3}\, ds_4^2$ and 
\be
d\hat s_7^2 = \lambda^{-2/3}\, ds_6^2 + \lambda^{4/3}\, (dz + {\cal A})^2\ .
\label{d7met}
\ee
Using the fact that if $d\hat s_{{\sst D}+1}^2 = c^{-1}\,ds_{\sst D}^2
+ c^2\, (dz+{\cal A})^2$, the vielbein components $\hat R_{ab}$ of the
Ricci tensor of $d\hat s_{{\sst D}+1}^2$ are related to the vielbein
components $R_{ij}$ of $ds_{\sst D}^2$ by
\be
\hat R_{ij} = c\,R_{ij} -\ft12 c^4\, {\cal F}_{ik}\, {\cal F}_j{}^k\ ,\qquad
\hat R_{zz} =\ft14 c^4 {\cal F}_{ij}\, {\cal F}^{ij}\ ,\qquad \hat R_{iz}
= \ft12 c^{5/2}\, \nabla^j{\cal F}_{ij}\ , \label{oneill}
\ee
we can easily see that the seven-dimensional metric $d\hat s_7^2$ in
(\ref{d11sol}) is Einstein (regardless of whether or not ${\cal M}_6$ is 
Einstein). We find
\[
{\hat F}_{\mu\nu\rho\sigma} =
6{\hat m}{\hat\epsilon}_{\mu\nu\rho\sigma},
\]
\[ 
{\hat R}_{\mu\nu} = -12{\hat m}^2\, {\hat g}_{\mu\nu}\,
\]
\be
{\hat R}_{ab} = 6{\hat m}^2 {\hat g}_{ab},\label{d7ein}
\ee 
where ${\hat m}={\lambda}^{1/3}m$ and ${\hat
\epsilon}_{\mu\nu\rho\sigma}$ is the Levi-Civita tensor on $AdS_4$ in
the rescaled metric. In fact this lifting to $D=11$ of the class of
solutions (\ref{d10sol}) in $D=10$ gives the Freund-Rubin
class of solutions \cite{Freund} of eleven-dimensional supergravity,
in which the metric is written as a direct sum of a four-dimensional
anti de-Sitter spacetime and a seven-dimensional Einstein space ${\cal
M}_7$.  To be more precise, one obtains by this means all Freund-Rubin
solutions where ${\cal M}_7$ can be written as a $U(1)$ bundle over
some $6$-manifold ${\cal M}_6$ \cite{Nilssonpope}, which satisfies the
conditions given above.  We shall refer to solutions with $m=+\mu$ and
$m=-\mu$ as left-handed or right-handed respectively, since the sign
dictates the orientation of ${\cal M}_{7}$.

     The period $\Delta z$ of of the coordinate $z$ on the compactifying
circle is not arbitrary, and is determined by the harmonic form $K_{mn}$ on
${\cal M}_6$. Specifically, the period $\Delta z$ is given by
\be
\Delta z= \int {\cal F}\ ,
\ee
or an integer fraction of this, where the integral is taken over
$2$-cycles in ${\cal M}_6$.  (If there is more than one $2$-cycle, the
periods determined by these integrals must be compatible, in order to
have a well-defined solution.)
  
     As we mentioned previously, a simple class of spaces ${\cal M}_6$ that
satisfy the necessary conditions are Einstein-K\"ahler spaces, with
$K_{mn}$ taken to be the K\"ahler form.  One such example is provided
by $CP^3$, endowed with the Fubini-Study metric.  In this case, the
seven-dimensional space ${\cal M}_7$, obtained as a $U(1)$ bundle over
$CP^3$, is the seven-sphere.  In fact the metric $d\hat s_7^2$ in this
example is precisely the standard $SO(8)$-invariant ``round''
seven-sphere.  We shall discuss this example in more detail in section
\ref{round}.  Another example with the same topology is obtained by
taking ${\cal M}_6$ to have a homogeneously ``squashed'' $CP^3$
metric, with isometry group $SO(5)$ rather than the $SU(4)$ isometry
of the standard Fubini-Study metric. In this case, by choosing the
squashing parameter appropriately, one obtains a solution of the type
IIA theory that lifts back \cite{Nilssonpope} to the squashed
seven-sphere solution \cite{Awadaduffpope,Duffnilssonpope2} of $D=11$
supergravity.  Interestingly enough, the $CP^3$ metric in this case is
neither Einstein nor K\"ahler, although it is still Hermitean
\cite{Nilssonpope}. See section \ref{squashed}.
 
     As discussed in section \ref{further}, further interesting
examples of ${\cal M}_7$ spaces that are $U(1)$ fibrations can be
obtained by taking ${\cal M}_6$ to be $CP^2 \times S^2$, $S^2\times
S^2\times S^2$, or the flag manifold $SU(3)/T^2$.  These give rise to
solutions of the Type $IIA$ theory whose liftings to $D=11$ describe
compactifications where the ${\cal M}_7$ are $M^{pqr}$, $Q^{pqr}$ or
$N^{pqr}$ spaces \cite{Nilssonpope}.  In the first two cases, the
${\cal M}_7$ spaces are obtained as $U(1)$ bundles characterised by
their winding numbers with respect to the various K\"ahler structures
in the $CP^2$ or $S^2$ subspaces.

     Let us consider the example where ${\cal M}_6$ is $CP^3$, with
its Fubini-Study metric, in more detail.  Since the metric is
Einstein-K\"ahler, with $R_{mn}=8 m^2\, g_{mn}$, it follows that the
Ricci form is given by $\rho= 8 m^2\, J$, and thus the period of the
compactifying coordinate $z$ is given in terms of the first Chern
class $c_1=\ft1{2\pi}\int\rho$ by
\be
\Delta z = \fft{\pi\,\mu\, c_1}{2\lambda\, m^2}\ .
\ee
Now the period $\Delta z$ is fixed once and for all, when the dimensional 
reduction from eleven dimensions is performed, and we may, without loss 
of generality, choose $\Delta z$ to have a convenient value.  Let us take 
$\Delta z= 2\pi L$ where L is an arbitrary length scale.  From the fact that
the  first Chern class for $CP^3$ is 
$c_1=4$, it then follows that $\mu =\lambda Lm^2$.  Combined with the 
relation $m^2=\mu^2$ in (\ref{d10sol}), we find that in this 
$CP^3$ solution the constant $m$ is given by 
\be
m=\fft{1}{L \lambda}\ .
\label{m}
\ee 
As a consistency check we note that $\hat
m=\lambda^{1/3}m=1/R_{11}¥$, where $R_{11}¥=L\lambda^{2/3}$ is
the radius of the eleventh dimension, as we would expect. This means
that the R-R fields in (\ref{d10sol}) are actually $O(\lambda^{-2})$

Although solutions of Type $IIA$ supergravity, we have no conformal
field theory argument to prove that the configurations
(\ref{d10sol}) are solutions of the full string theory.  There do
exist conformal field theories with $AdS$ vacua
\cite{Antoniadis1,Antoniadis2} but they involve only Neveu-Schwarz
fields.  However, (\ref{d10sol}) involves R-R field strengths which
one does not know how to incorporate into conformal field theory
except by expanding in powers of the R-R field, but this we cannot
do because they are $O(\lambda^{-2})$ \footnote{We are grateful to
E. Witten for correspondence on this point.}.
 
\section{The round seven-sphere}
\label{round}

     Considered as a compactification of $D=11$ supergravity, the round
$S^7$ yields a four dimensional $AdS$ spacetime with $N=8$
supersymmetry and $SO(8)$ gauge symmetry, for either orientation of
$S^7$.  The Kaluza-Klein mass spectrum therefore falls into $SO(8)$
$N=8$ supermultiplets.  In particular, the massless sector is
described by gauged $N=8$ supergravity \cite{Klein}.  Since $S^7$ is a
$U(1)$ bundle over $CP^3$ the same field configuration is also a
solution of $D=10$ Type $IIA$ supergravity.  However, the resulting
vacuum has only $SU(4) \times U(1)$ symmetry and either $N=6$ or $N=0$
supersymmetry depending on the orientation of the $S^7$.  The reason
for the discrepancy is that the modes charged under the $U(1)$ are
associated with the Kaluza-Klein reduction from $D=11$ to $D=10$ and
are hence absent from the Type $IIA$ spectrum originating from the
massless Type $IIA$ supergravity.  In other words, they are Dirichlet
$0$-branes and hence absent from the perturbative string
spectrum. There is thus more non-perturbative gauge symmetry and
supersymmetry than perturbative. The right-handed orientation is
especially interesting because the perturbative theory has no
supersymmetry at all! See Table \ref{massless} (where we are using the
notation of \cite{slan} for $SU(4)$ representations). It is interesting 
to note that the $D=4$ massless states in the left-handed vacuum 
originate from the $n=0$ massless level and $n=1,2$ massive levels in 
$D=10$, whereas in the right-handed vacuum they originate from 
$n=0,1,2,3,4$ levels. 

\begin{table}
\begin{center}
\begin{tabular}{|c|c|l|l|}\hline
Spin & $SO(8)$ reps & Left $SU(4) \times U(1)$ reps & Right $SU(4) \times U(1)$
reps\\  \hline\hline
$2$  & $1$    & $1_0$ & $1_0$ \\
$\ft32$& $8_s$  & $6_0+1_2+1_{-2}$ & $4_1+{\bar 4}_{-1}$ \\
$1$  & $28$   & $1_0+15_0+6_2+6_{-2}$ & $1_0+15_0+6_2+6_{-2}$ \\ 
$\ft12$& $56_s$ & $6_0+10_0+{\bar {10}}_0+15_2+15_{-2}$ & $4_1+{\bar
4}_{-1}+20_1+{\bar {20}}_{-1}+4_{-3}+{\bar 4}_{3}$ \\ $0^+$& $35_v$ &
$15_0+10_{-2}+{\bar {10}}_2$ & $15_0+10_{-2}+{\bar {10}}_2$ \\ $0^-$& $35_c$ &
$15_0+10_{2}+{\bar {10}}_{-2}$ & $1_0+{20'}_0+6_2+6_{-2}+1_4+1_{-4}$ \\ \hline
\end{tabular}
\end{center}
\caption{The massless multiplet under $SO(8) \rightarrow SU(4) \times U(1)$}
\label{massless}
\end{table}

     Similar remarks apply to the massive spectrum.  By way of
examples, we discuss the decomposition of the first two massive
levels, for both the left-handed $N=6$ vacuum in Appendix A, and the
right-handed $N=0$ vacuum in Appendix B.

\section{The squashed seven-sphere}
\label{squashed}

The squashed $S^7$ solution of $D=11$ supergravity yields an
$M$-theory vacuum with $SO(5) \times SU(2)$ and $N=1$ or $N=0$ for
either left or right orientation
\cite{Awadaduffpope,Duffnilssonpope2}. Again it corresponds to a
$U(1)$ bundle over a squashed $CP^3$ and may also be interpreted
\cite{Nilssonpope} as a solution of Type $IIA$ supergravity but with
only $SO(5) \times U(1)$.  Since the single massless gravitino of the
left-squashed solution is a singlet under $U(1)$, the perturbative
vacua have the same supersymmetry as the non-perturbative.

\section{$M(3,2)$}
\label{further}
 
     The $M^{pqr}$ spaces can be described very simply as $U(1)$
bundles over $CP^2\times S^2$.  The most natural way to classify them
is in the $M(m,n)$ notation of \cite{papo}, where $m$ and $n$ are the
winding numbers of the $U(1)$ fibres over $CP^2$ and $S^2$
respectively.  In the notation of \cite{Fre}, the space $M^{pqr}$
corresponds to $M(m,n)$ with $p=m/r$ and $q=n/r$, where $r$ is the
greatest common divisor of $m$ and $n$.  For each pair of integers $m$
and $n$, there is an Einstein metric on the 7-dimensional bundle
space, where the scale sizes of the standard metrics on $CP^2$ and
$S^2$ are chosen appropriately \cite{papo}.  In the case $M(3,2)$, the
space admits two Killing spinors; when $2m\ne 3n$, there are no
Killing spinors.

   This example is particularly interesting for two reasons. First, it
is a special case of the $M^{pqr}$ spaces constructed by Witten
\cite{Witten} displaying $SU(3) \times SU(2) \times U(1)$
symmetry. This symmetry is the same even when considered as a solution
of Type $IIA$ supergravity. Secondly, as an $M$-theory vacuum the
right-handed orientation has $N=0$ \cite{Klein} and the left-handed
orientation has $N=2$ \cite{Fre}, but since both gravitinos are
charged under the $U(1)$, the perturbative theory has no
supersymmetry.  It fact, the perturbative theory is especially
bizarre: it has no fermions whatsoever! This because $M^{pqr}$ is a
$U(1)$ bundle over $CP^2 \times S^2$ and although it has a spin
structure, $CP^2 \times S^2$ does not. Only spinors with appropriate
charges under the R-R $2$-form field strength on $CP^2$ can be
defined. These are $spin^c$ spinors or generalized spinors
\cite{Hawkingpope}.  Hence the perturbative spectrum of such a Type
$IIA$ vacuum is entirely bosonic: all the fermions are Dirichlet
$0$-branes!

\section{Perturbative versus non-perturbative}
\label{versus}

The states that are charged under the R-R $U(1)$ are Dirichlet
$0$-branes whose masses are of order $1/\lambda$ and which therefore
become very heavy for weak coupling of the $D=10$ Type $IIA$ string.
On the other hand they belong to the same supermultiplet as the
neutral states and should therefore be degenerate in mass with them
(in the sense of $AdS$ supersymmetry).  Consistency therefore demands
that {\it all} the four-dimensional massive states have masses of
order $1/\lambda$, whether Dirichlet $0$-branes or not, and that the
entire massive Kaluza-Klein spectrum is, in this sense,
non-perturbative.  This is indeed the case because the massive
spectrum associated with the compactification from $D=10$ to $D=4$ on
$CP^{3}$ is governed by the parameter $m$ which, from (\ref{m}) does
indeed go like $1/\lambda$.  Note, however, that the massless sector
remains massless for any value of $\lambda$ and so it still makes
sense to talk about ``perturbative'' supersymmetry whose gravitinos
are neutral under the $U(1)$ and the extra ``non-perturbative''
supersymmetry whose gravitinos are Dirichlet $0$-branes.  Similar
remarks apply to the gauge symmetry and the massless gauge bosons.

The $AdS$ vacua of Type $IIA$ supergravity considered in this paper
have spacetime cosmological constant $m^2 \sim 1/\lambda^{2}$ and are,
of course, very different from the Minkowski vacua.  Note also that
the volume $V$ of ${\cal M}_{6}$ scales like $m^{-6} \sim
\lambda^{6}$.  This has a strange consequence. From the four
dimensional point of view, the string coupling $\lambda_{4}$ is
related to the ten-dimensional string coupling $\lambda$ appearing in
(\ref{d10lag}) by
\be
\frac{1}{\lambda_{4}^{2}}=\frac{V}{L^{6}\lambda^{2}} 
\ee
In Minkowski space string theory, the volume $V$ is given by the vev
of a massless modulus field and its value is independent of the string
coupling which is determined by the vev of the dilaton $\phi$.
Consequently, weak coupling in $D=10$ means weak coupling in $D=4$.
In $AdS$, however, the volume field belongs to the massive sector
\cite{Klein}. In the case of the $CP^{3}$ for example, it is the
singlet scalar in the second massive level that comes from the singlet
in the round $S^{7}$ supermultiplet with spins
$\{2,\ft32,1,\ft12,0^+,0^-\}$ transforming as
$\{35_v,56_{s}¥+224_{vs}¥,28+350+567_{v}¥,8_{s}¥+160_s
+672_{vc} +840_{s}¥, 1+294_{v}¥+300,35_{s}¥+840'_{s}¥\}$,
as discussed in Appendices A and B.  Moreover, the vev of this scalar
is not a free parameter but is itself fixed by the string
coupling. From (\ref{d10sol}) and (\ref{m}), we see that
\be
V\sim m^{-6}=L^{6}\lambda^{6}
\ee
Thus we reach the bizarre conclusion that
\be
\lambda_{4}\sim \frac{1}{\lambda^{2}}
\ee
and the four-dimensional string coupling grows as the ten-dimensional 
string coupling shrinks! Similar remarks apply to the non-abelian 
gauge coupling constant $e$ ($SO(8)$ in the case of the round $S^{7}$). 
Since one of the gauge bosons is just the $U(1)$ gauge field appearing 
in (\ref{d10lag}), we have
\be
e^{2}=\frac{L^{6}}{V}\sim \frac{1}{\lambda^{6}¥} 
\ee

\section{Conclusions}
\label{conclusions}

We have seen how supersymmetry can appear very differently when viewed 
from the perturbative Type $IIA$ string perspective (vacua 
corresponding to solutions of massless Type $IIA$ supergravity whose 
spectrum is neutral under the R-R charge) and 
the non-perturbative $M$-theory perspective (vacua corresponding  
to solutions of $D=11$ supergravity whose spectrum includes Dirichlet 
$0$-branes). A summary of perturbative versus 
non-perturbative symmetries is given in Table \ref{symmetries}. In 
particular, the non-perturbative vacuum may have unbroken 
supersymmetry even when the perturbative vacuum has none.
\begin{table}
\begin{center}
\begin{tabular}{|c|cc|cc|}\hline
Compactification& &Perturbative Type $IIA$& &Nonperturbative M-theory\\ 
\hline\hline
Left round $S^7$ & $N=6$ & $SU(4) \times U(1)$ & $N=8$ & $SO(8)$\\
Right round $S^7$ & $N=0$ & $SU(4) \times U(1)$ & $N=8$ & $SO(8)$\\
Left squashed $S^7$ & $N=1$ & $SO(5) \times U(1)$ & $N=1$ &  $SO(5) \times
SU(2)$\\ 
Right squashed $S^7$ & $N=0$ & $SO(5) \times U(1)$ & $N=0$ &  $SO(5) \times
SU(2)$\\
Left $M(3,2)$ & $N=0$ & $SU(3) \times SU(2) \times U(1)$ & $N=2$ & $SU(3) 
\times SU(2) \times U(1)$\\
Right $M(3,2)$ & $N=0$ & $SU(3) \times SU(2) \times U(1)$ & $N=0$ & $SU(3) 
\times SU(2) \times U(1)$\\ \hline
\end{tabular}
\end{center}
\caption{Perturbative versus non-perturbative symmetries}
\label{symmetries}
\end{table}

We cannot resist asking whether this could be a model of the real
world in which you can have your supersymmetry and eat it too
\footnote{A scheme in which you can have all the benefits of unbroken
supersymmetry while appearing to inhabit a non-supersymmetric world
has also been proposed by Witten \cite{Wittenworld} but his mechanism
is very different from ours. In particular, our vacua necessarily have
non-vanishing cosmological constant unless cancelled by fermion
condensates \cite{Orzalesi}.}. The problem with such a scenario, of
course, is that God does not do perturbation theory and presumably an
experimentalist would measure God's real world and not what a
perturbative string theorist thinks is the real world. Unless, for
some unknown reason, the experimentalist's apparatus is so primitive
as to be unable to detect Ramond-Ramond charge in which case he or she
would conclude that the world has no unbroken supersymmetry.

\section{Acknowledgements}

We have enjoyed useful conversations with Karim Benakli, Jacques 
Distler, Glennys Farrar, Hermann Verlinde and Edward Witten.

\appendix
\section{The first two massive levels on $S^7$: $N=6$ decomposition}

Let us consider the first massive level \cite{Klein} on $S^7$.  Under 
the $N=8 \rightarrow N=6$, $SO(8) \rightarrow SU(4) \times U(1)$ 
decomposition of the left-handed $S^7$, it yields the decomposition 
given in Table \ref{firstmassiveleft}.  Note that there are no $U(1)$ 
singlets so all these states are Dirichlet $0$-branes, absent from the 
Type $IIA$ Kaluza-Klein spectrum in going from $D=10$ to $D=4$.  This 
is a general feature of the odd numbered levels.

Next we look at the second massive level in Table 
\ref{secondmassiveleft}.  Since the numbers of representations in the 
decompositions rapidly become large, we shall just present the subset 
of states that carry no Ramond-Ramond charges and which survive in the 
truncation to the Type $IIA$ spectrum forming $N=6$ supermultiplets.

\begin{table}
\begin{center}
\begin{tabular}{|c|c|l|}\hline
Spin & $SO(8)$ reps & Left $SU(4) \times U(1)$ reps \\  
\hline\hline
$2$  & $8_v$&$4_{-1} +\bar 4_{1}$\\
$\ft32$& $8_c$&$4_1+\bar 4_{-1}$\\
 ~    & $56_{c}¥$&$4_1 +\bar 4_{-1} +20_1+\bar{20}_1 +4_{-3} +\bar 4_3$\\
$1$  & $56_{v}¥$&$4_{-1}+\bar 4_1 + 20_{-1} +\bar{20}_1 + 4_3 +
\bar 4_{-3}$\\ 
 ~    & $160_{v}¥$&$4_{-1}+\bar 4_1 + 20_{-1} +\bar{20}_1 + 20_3 + 
\bar{20}_{-3} + 36_{-1} +\bar{36}_1$\\
$\ft12$& $160_{c}¥$ &$4_{1}+\bar 4_{-1} + 20_{1} +
\bar{20}_{-1} + 20_{-3} + \bar{20}_{3} + 36_{1} +\bar{36}_{-1}$\\ 
 ~    & $224_{vc}¥$&$20_{1} +\bar{20}_{-1} +20''_1 +\bar{20}''_{-1}
+ 36_1+\bar{36}_{-1} + 36_{-3} +\bar{36}_{3}$\\
$0^+$& $112_{v}¥$ &$20''_3 + \bar{20}''_{-3} + 36_{-1} +\bar{36}_1$\\
$0^-$& $224_{cv}¥$ &$20_{-1} +\bar{20}_{1} +20''_{-1} 
+\bar{20}''_{1}+ 36_{-1}+\bar{36}_{1} + 36_{3} +\bar{36}_{-3}$\\ \hline
\end{tabular}
\end{center}
\caption{The first massive level under $SO(8) \rightarrow SU(4) \times U(1)$}
\label{firstmassiveleft}
\begin{center}
\begin{tabular}{|c|c|l|}\hline
Spin & $SO(8)$ reps & Left $SU(4) \times U(1)$ reps \\  
\hline\hline
$2$& $35_v$&$15_0 +\cdots\ $\\
$\ft32$&$56_s$&$6_0+10_0+\bar{10}_0+\cdots\ $\\
 ~&$224_{vs}$&$10_0 + \bar{10}_0 + 64_0 +\cdots$ \\
$1$& $28$&$1_0 +15_0 +\cdots\ $\\
 ~&$350$&$15_0+15_0 +20'_0 +45_0 +\bar{45}_0 +\cdots\ $\\
 ~&$567_v$&$15_0 + 45_0 + \bar{45}_0 + 84_0+\cdots\ $\\
$\ft12$&$8_s$&$6_0 +\cdots \ $\\
 ~&$160_s$&$6_0 +6_0 +64_0 +\cdots\ $\\
 ~&$672_{vc}$&$64_0 +70_0+\bar{70}_0+\cdots     \ $\\
 ~&$840_s$&$6_0 +10_0 +\bar{10}_0+ 64_0 +64_0+ 70_0 + \bar{70}_0 +\cdots\ $\\
$0^+$&$1$&$1_0$\\
 ~&$294_v$&$0+\cdots \ $\\
 ~&$300$& $1_0 + 15_0 +20'_0+84_0 +\cdots \ $\\
$0^-$& $35_s$&$1_0 + 20'_0 +\cdots\ $\\
 ~&$840'_s $&$20'_0 +35_0 + \bar{35}_0 + 45_0 +\bar{45}_0 + 84_0 +\cdots \ $
\\ \hline
\end{tabular}
\end{center}
\caption{The second massive level under $SO(8) \rightarrow SU(4) \times U(1)$}
\label{secondmassiveleft}
\end{table}

\section {The first two massive levels on $S^7$: $N=0$ decomposition}

Again let us first consider the first massive level on $S^7$.  Under the 
$N=8 \rightarrow N=0$, $SO(8) \rightarrow SU(4) \times U(1)$ decomposition 
of the right-handed $S^7$, it yields the decomposition given in Table 
\ref{firstmassiveright}.  Note that the surviving singlets are all fermions. 
This is a general feature of the odd numbered levels. 

Next we look at the second massive level in Table
\ref{secondmassiveright}.  Note that the surviving singlets are all
bosons. This is a general feature of all even numbered levels. Keeping
just the $U(1)$ singlets thus yields a massive spectrum of bosons and
fermions but which do not form supermultiplets. The supermultiplet
structure of the spectrum becomes apparent only when the Dirichlet
$0$-branes are incorporated.

The coupling constants appearing in the truncated tree-level Lagrangian would 
take on their supersymmetric values but this would not persist at 
the quantum level unless the Dirichlet $0$-branes were re-introduced. 

\begin{table}
\begin{center}
\begin{tabular}{|c|c|l|}\hline
Spin & $SO(8)$ reps & Right $SU(4) \times U(1)$ reps \\  
\hline\hline
$2$  & $8_v$&$4_{-1} +\bar 4_{1}$\\
$\ft32$& $8_c$&$1_{2}¥+1_{-2}¥ +6_{0}¥$\\
 ~    & $56_{c}¥$&$6_0 + 10_0 + \bar{10}_0 + 15_2 + 15_{-2}$\\
$1$  & $56_{v}¥$&$4_{-1}+\bar 4_1 + 20_{-1} +\bar{20}_1 + 4_3 +
\bar 4_{-3}$\\ 
 ~    & $160_{v}¥$&$4_{-1}+\bar 4_1 + 20_{-1} +\bar{20}_1 + 20_3 + 
\bar{20}_{-3} + 36_{-1} +\bar{36}_1$\\
$\ft12$& $160_{c}¥$ &$1_2+ 1_{-2} + 6_0 + 6_0 + 6_4 + 
6_{-4} + 15_2 + 15_{-2}$\\ 
 ~    & $224_{vc}¥$&$10_0 +\bar{10}_0 + 64_0 + 15_2 + 15_{-2} + 
45_{-2} + \bar{45}_2 + 10_{-4} + \bar{10}_4$\\
$0^+$& $112_{v}¥$ &$20''_3 + \bar{20}''_{-3} + 36_{-1} +\bar{36}_1$\\
$0^-$& $224_{cv}¥$ &$4_{-1} +\bar 4_1 + 4_3 + \bar 
4_{-3} + 4_{-5} + \bar 4_{5} + 20_{-1} + \bar{20}_1 + 20_3 + \bar{20}_{-3} + 
60_{-1} + \bar{60}_1 $\\ \hline
\end{tabular}
\end{center}
\caption{The first massive level under $SO(8) \rightarrow SU(4) \times U(1)$}
\label{firstmassiveright}
\begin{center}
\begin{tabular}{|c|c|l|}\hline
Spin & $SO(8)$ reps & Right $SU(4) \times U(1)$ reps \\  
\hline\hline
$2$& $35_v$&$15_0 +\cdots\ $\\
$\ft32$&$56_s$&$0+\cdots$\\
 ~&$224_{vs}$&$0+\cdots$ \\
$1$& $28$&$1_0 +15_0 +\cdots\ $\\
 ~&$350$&$15_0+15_0 +20'_0 +45_0 +\bar{45}_0 +\cdots\ $\\
 ~&$567_v$&$15_0 + 45_0 + \bar{45}_0 + 84_0+\cdots\ $\\
$\ft12$&$8_s$&$0+\cdots$\\
 ~&$160_s$&$0+\cdots$\\
 ~&$672_{vc}$&$0+\cdots$\\
 ~&$840_s$&$0+\cdots$\\
$0^+$&$1$&$1_0$\\
 ~&$294_v$&$0+\cdots \ $\\
 ~&$300$& $1_0 + 15_0 +20'_0+84_0 +\cdots \ $\\
$0^-$& $35_s$&$15_{0}+\cdots \ $\\
 ~&$840'_s $&$10_{0}¥+45_{0}¥+{\bar 45}_{0}¥ +175_{0}¥$\\ \hline
\end{tabular}
\end{center}
\caption{The second massive level under $SO(8) \rightarrow SU(4) \times U(1)$}
\label{secondmassiveright}
\end{table}


\newpage
\begin{thebibliography}{99}

\bibitem{Howe}
M. J. Duff, P.~Howe, T.~Inami and K. S. Stelle,
\newblock {\sl Superstrings in $D = 10$ from supermembranes in $D = 11$},
\newblock Phys. Lett. {\bf B191} (1987) 70.

\bibitem{Luduality}
M. J. Duff and J. X. Lu,
\newblock {\sl Duality rotations in membrane theory},
\newblock Nucl.Phys. {\bf B347} (1990) 394.

\bibitem{Hulltownsend}
C. M. Hull and P. K. Townsend,
\newblock {\sl Unity of superstring dualities},
\newblock Nucl. Phys. {\bf B438} (1995) 109.

\bibitem{Townsendeleven}
P. K. Townsend,
\newblock {\sl The eleven-dimensional supermembrane revisited},
\newblock Phys.Lett.{\bf B350} (1995) 184.

\bibitem{Wittenvarious}
E.~Witten,
\newblock {\sl String theory dynamics in various dimensions},
\newblock Nucl. Phys. {\bf B443} (1995) 85.

\bibitem{Duffliuminasian}
M. J. Duff, J.~T. Liu and R.~Minasian,
\newblock {\sl Eleven-dimensional origin of string/string duality: A one 
loop test},
\newblock Nucl.Phys. {\bf B452} (1995) 261.

\bibitem{Becker1}
K.~Becker, M.~Becker and A.~Strominger,
\newblock {\sl Fivebranes, membranes and nonperturbative string theory},
\newblock Nucl.Phys. {\bf B456} (1995) 130.

\bibitem{Schwarzpower}
J. H. Schwarz,
\newblock {\sl The power of $M$-theory},
\newblock Phys. Lett. {\bf B360} (1995) 13.

\bibitem{Horava}
P. Horava and E. Witten,
\newblock {\sl Heterotic and Type $I$ string dynamics from eleven 
dimensions},
\newblock Nucl. Phys. {\bf B460} (1996) 506.

\bibitem{TownsendM}
P. Townsend,
\newblock {\sl $D$-Branes From $M$-Branes},
\newblock Phys. Lett. {\bf B373} (1996) 68.

\bibitem{Aharony}
O. Aharony, J. Sonnenschein and S. Yankielowicz,
\newblock {\sl Interactions of strings and $D$-branes from $M$-theory},
\newblock  Nucl. Phys. {\bf B474} (1996) 309.

\bibitem{DuffM}
M. J. Duff,
\newblock {\sl  $M$-theory (the theory formerly known as strings)},
\newblock  I. J. M. P {\bf A11} (1996) 5623. 

\bibitem{BanksM}
T. Banks, W. Fischler, S. H. Shenker and L. Susskind, 
\newblock {\sl $M$ theory as a matrix model: a conjecture},
\newblock Phys. Rev. {\bf D55} (1997) 5112.

\bibitem{Polchinski}
J.~Polchinski,
\newblock {\sl Dirichlet-branes and Ramond-Ramond charges},
\newblock Phys. Rev. Lett. {\bf 75} (1995) 4724.

\bibitem{Klein}
M.~J. Duff, B.~E.~W. Nilsson and C.~N. Pope,
\newblock {\sl {K}aluza-{K}lein supergravity},
\newblock Phys. Rep. {\bf 130} (1986) 1.

\bibitem{Cremmer}
E.~Cremmer, B.~Julia and J.~Scherk,
\newblock {\sl Supergravity theory in $11$ dimensions},
\newblock Phys. Lett. {\bf B76} (1978) 409.

\bibitem{Duffnilssonpope3}
 M. J. Duff, B. E. W. Nilsson and C. N. Pope,
\newblock {\sl Compactification of $D=11$ supergravity on $K3 \times T^{3}$},
\newblock Phys. Lett. {\bf B129} (1983) 39.

\bibitem{Ovrut}
A. Lucas, B. A. Ovrut and D. Waldram,
\newblock {\sl Stabilizing dilaton and moduli vacua in string and 
$M$-theory},
\newblock {\tt hep-th/9611204}.

\bibitem{Nilssonpope}
B. E. W. Nilsson and C. N. Pope,
\newblock {\sl Hopf fibration of eleven-dimensional supergravity},
\newblock Class. Quantum Grav. {\bf 1} (1984) 499.

\bibitem{Duffstelle}
M. J. Duff and K.~Stelle,
\newblock {\sl Multimembrane solutions of $d = 11$ supergravity},
\newblock Phys.Lett. {\bf B253} (1991) 113.

\bibitem{Gibbonstownsend} 
G. W. Gibbons and P. K. Townsend,
\newblock{\sl Vacuum interpolation in supergravity via super $p$-branes},
\newblock Phys. Rev. Lett. {\bf 71} (1993) 3754.

\bibitem{Gibbonsdufftownsend}
M. J. Duff, G.~W. Gibbons and P.~K. Townsend,
\newblock {\sl Macroscopic superstrings as interpolating solitons},
\newblock Phys. Lett. {\bf  B332} (1994) 321. 

\bibitem{Green}
M. B. Green and J. S. Schwarz,
\newblock {\sl Covariant description of superstrings},
\newblock Phys. Lett. {\bf  B136} (1984) 367. 

\bibitem{Bergshoeff}
E.~Bergshoeff, E.~Sezgin and P.~Townsend,
\newblock {\sl Supermembranes and eleven-dimensional supergravity}
\newblock Phys.Lett. {\bf B209} (1988) 451.

\bibitem{Campbellwest}
I. G. Campbell and P. C. West, 
\newblock {\sl $N=2$ $D=10$ nonchiral supergravity and its spontaneous 
compactification},
\newblock Nucl. Phys.{\bf B243} (1984) 112.

\bibitem{Huq}
M. Huq and M. A. Namazie, 
\newblock {\sl Kaluza-Klein supergravity in ten dimensions},
\newblock Class. Quant. Grav. {\bf 2} (1985) 293.

\bibitem{Freund}
P. G. O. Freund and M. A. Rubin,
\newblock {\sl Dynamics of dimensional reduction},
\newblock Phys. Lett. {\bf B97} (1980) 233.

\bibitem{slan}
R. Slansky,
\newblock {\sl Group theory for unified model building},
\newblock Phys. Rep. {\bf 79} (1981) 1.

\bibitem{Awadaduffpope} M. A. Awada, M. J. Duff and C. N. Pope, 
\newblock {\sl $N = 8$ Supergravity Breaks Down to $N = 1$}, 
\newblock Phys. Rev. Lett. {\bf 50}, 294 (1983).  

\bibitem{Duffnilssonpope2} M. J. Duff, B. E. W. Nilsson and C. N. Pope, 
\newblock{\sl Spontaneous Supersymmetry Breaking by the Squashed 
Seven-Sphere}, 
\newblock Phys. Rev. Lett. {\bf 50}, 2043 (1983).  

\bibitem{papo}
D.N. Page and C. N. Pope,
\newblock {\sl Stability analysis of compactifications of $D=11$ 
supergravity with $SU(3) \times SU(2) \times U(1)$ symmetry},
\newblock Phys. Lett. {\bf B145} (1984) 337.

\bibitem{Witten}
E. Witten,
\newblock {\sl Search for a realistic Kaluza Klein theory},
\newblock Nucl. Phys. {\bf B186} (1981) 412.

\bibitem{Fre}
L. Castellani, R. D'Auria and P. Fre, 
\newblock {\sl $SU(3) \times SU(2) \times U(1)$ from $D=11$ supergravity},
\newblock Nucl. Phys. {\bf B239} (1984) 610.

\bibitem{Hawkingpope}
S. W. Hawking and C. N. Pope,
\newblock {\sl Generalized spin structures in quantum gravity},
\newblock Phys. Lett. {\bf B73} (1978) 42.

\bibitem{Antoniadis1}
I. Antoniadis, C. Bachas and A. Sagnotti, 
\newblock {\sl Gauged supergravity vacua in string theory},
\newblock Phys. Lett. {\bf B235} (1990) 255.

\bibitem{Antoniadis2}
I. Antoniadis, S. Ferrara and C. Kounnas, 
\newblock {\sl Exact supersymmetric string solutions in curved 
gravitational backgrounds},
\newblock Nucl. Phys.{ \bf B421} (1994) 343.

\bibitem{Wittenworld}
E. Witten, 
\newblock {\sl Strong coupling and the cosmological constant}, 
\newblock Mod. Phys. Lett. {\bf A10} (1995) 2153.

\bibitem{Orzalesi}
M. J. Duff and C. Orzalesi, 
\newblock {\sl The cosmological constant in
spontaneously compactified $D = 11$ supergravity}, 
\newblock Phys. Lett. {\bf B122} (1983) 37. 


\end{thebibliography}
\end{document}